\documentclass[preprint,a4paper,nofootinbib]{revtex4-1}
\usepackage[utf8]{inputenc}
\usepackage{textcomp}
\usepackage{gensymb}
\usepackage{amsmath}
\usepackage{amssymb}
\usepackage{xspace}
\usepackage{nicefrac}
\usepackage{xcolor}
\usepackage{graphicx}
\usepackage{soul}
\usepackage{hyperref}
\usepackage{subfig}
\usepackage{multirow}

\usepackage[titletoc,toc,page]{appendix}

\begin{document}
\title{Probing UHECR production in Centaurus A using secondary neutrinos and gamma-rays}
\date{\today}
\author{Cainã de Oliveira}
\email{caina.oliveira@ifsc.usp.br}
\author{Vitor de Souza}
\email{vitor@ifsc.usp.br}
\affiliation{Instituto de F\'isica de S\~ao Carlos, Universidade de S\~ao Paulo, Av. Trabalhador S\~ao-carlense 400, S\~ao Carlos, Brasil.}

\begin{abstract}

  In this paper the production of neutrinos and photons by ultra high energy cosmic rays (UHECR) interacting with the extragalactic background radiation is studied. Centaurus A is assumed as the prime source of UHECR and the possibility to identify this source by detecting the secondary neutrinos and photons produced in the propagation of UHECR is investigated. Fifteen astrophysical models regarding three extragalactic magnetic fields (EGMF) and five composition abundances are simulated. The flux and arrival direction of neutrinos and photons are investigated. It is shown that the detection of a signal from Cen A with statistical significance is achievable by current observatories in a few years and by proposed experiments in the near future. The dependence of the results on the models is also presented.

\end{abstract}

\maketitle

\section{Introduction}

The unambiguous identification of an ultra high energy cosmic ray (UHECR) source would bring an enormous advance to astroparticle physics. Several unknowns related to the lack of knowledge of the source type, propagation effects and composition would be drastically reduced if one source is detected. The most famous candidate is Centaurus A (Cen A), the closest radio galaxy with a very rich morphology containing a radio jet extending over kiloparsecs and giant-lobes with kiloparsec size~\cite{burns1983inner, israel1998centaurus}. Several studies have shown that Cen A has the necessary conditions to accelerate particles at the highest energies~\cite{romero1996centaurus, dermer2009ultra, biermann2010ultra, biermann2012centaurus, wykes2013mass, wykes2018uhecr, matthews2018fornax, matthews2019ultrahigh}. Its relative size and proximity helps the detection and direct experimental evidences for acceleration of high energy particles have been published~\cite{hartman1999third, aharonian2009discovery, caccianiga2019anisotropies, hess2020resolving}.

A multimessenger approach has been advocated as the only way to give definitive answers to UHECR production. Neutral particles (neutrons, neutrinos and photons) are considered to be the smoking gun of the UHECR source since they do not deviate on the way to Earth. Intrinsic TeV gamma-rays have been detected from Cen A~\cite{hess2020resolving} however the connection to UHECR production depends on the unknown characteristics of the acceleration sites~\cite{liu2017particle}. The uncertainties about the environment of production do not allow one to infer if high energetic photons and neutrinos are produced as a consequence of UHECR acceleration.

Models for the emission of primary neutrinos as a consequence of proton acceleration in Cen A have already been proposed to describe the core and jet emission~\cite{cuoco2008ultrahigh,kachelriess2009high}~\footnote{Heavier nuclei acceleration would produce less neutrinos, therefore the neutrino flux prediction in these models represents an upper limit.}. The original models were optimized in 2008 and 2009 using the first years of data measured by the Pierre Auger Observatory. A large uncertainty was introduced in the original proposals in the normalization of the neutrino flux which was taken to be proportional to the UHECR flux around a certain radius centered in Cen A. Figure~\ref{fig:neutrino:intrinsic} shows an update of these original models to the last data release of the Pierre Auger Observatory, see appendix~\ref{app:flux} for details. The ignorance about the contribution of Cen A to the total flux of UHECR is shown as a shaded area that covers a range of contribution of Cen A to the total flux measured by the Pierre Auger Observatory between $1$ and $100\%$. The $90\%$ upper limits on the neutrino flux measured by the Pierre Auger Observatory and by the IceCube are also shown for comparison. The prediction of sensitivity of future experiments ARA~\cite{ARA2016}, ARIANNA~\cite{ARIANNA}, POEMMA~\cite{venters2019poemma}, IceCube\_Gen2~\cite{aartsen2020icecube} and GRAND~\cite{grand2018} are also shown. Note that the sensitivity of the future experiments are not corrected to the sensitivity in the direction of Cen A. They are shown here only for comparison of the overall flux. The primary flux of neutrinos for Cen A shown in this figure is of the same order of the flux predicted by cosmogenic neutrinos~\cite{batista2019cosmogenic, heinze2019new}. Future experiments will have sensitivity to detect intrinsic neutrinos from Cen A if: a) the models predicting the emission from the core are correct and b) Cen A contributes to more than roughly 50\% of the total UHECR flux measured by the Pierre Auger Observatory.

Given that the intrinsic flux of photons and neutrinos at Cen A is or can be soon detected, the next sections delve into a possibly richer question concerning UHECR. The calculations presented here investigate for the first time if photons and neutrinos produced on the way from Cen A to Earth by UHECR could also be detected by current or future experiments. An UHECR interacts with the photon field and produces, via photo-pion production and nuclear fragmentation, photons and neutrinos. The mechanisms of production are well known which leads to a very precise prediction within a certain astrophysical scenario. For this reason, the detection of secondary neutrinos and photons would correlate more clearly to UHECR rather than the intrinsic flux of photons and neutrinos.

In this paper, the arrival direction and flux of secondary photons and neutrinos are investigated for fifteen astrophysical scenarios. In the next section the astrophysical scenarios will be presented and discussed. In section~\ref{sec:result:neutrino} and~\ref{sec:result:photon} the results concerning flux and arrival direction of neutrinos and photons are respectively presented. Section~\ref{sec:conclusion} presents the conclusions.

\section{Astrophysical Scenarios for Secondary Neutrinos and Photons}
\label{sec:astro:scenarios}

The prediction of the flux and arrival direction of particles coming from Cen A depends significantly on the astrophysical scenarios taken into account. In this paper, fifteen astrophysical scenarios are considered representing three extragalactic magnetic field models and five composition abundances. For each scenario, the propagation of cosmic rays from Cen A to Earth is simulated and the numbers of secondary neutrinos and photons are calculated.

It is not the purpose of this study to build an astrophysical scenario compatible to current data. Instead, fifteen astrophysical published scenarios were chosen and the corresponding secondaries photons and neutrinos were simulated. The astrophysical scenarios used here were built based on very different approaches and assumptions. It is the purpose of this paper to use a wide variety of astrophysical scenarios to calculate the secondary neutrino and photon flux produced by UHECR. In this way, it will be possible to understand the dependencies of the calculated fluxes on the models and to point out the conclusions valid under all assumptions. In this section, the astrophysical scenarios and the propagation assumptions are explained.

\subsection{Composition abundance at the source}

The exact composition of particles injected by the source is as yet unknown. In this paper, the composition proposed in references~\cite{eichmann2019high, eichmann2018ultra, wykes2018uhecr, aloisio2014ultra, liu2012excess} are used. The fundamentals of each model are described below:

\begin{description}

\item[Composition 1 (C1)~\cite{eichmann2018ultra, eichmann2019high}] abundance of the elements in the solar system.
\item[Composition 2 (C2)~\cite{wykes2018uhecr}] model based on realistic stellar population estimation in Cen A. The mass entrained in the lobe is converted in a numerical fraction and, for simplicity, the isotopes are mapped by $N_{He} = M(^3He)/3 + M(^4He)/4$, $N_{N} = M(^{12}C)/12 + M(^{14}N)/14 + M(^{16}O)/16$, $N_{Si} = M(^{20}Ne)/20 + M(^{22}Ne)/22 + M(^{24}Mg)/24 + M(^{26}Mg)/26 + M(^{28}Si)/28$ and $N_{Fe} = M(^{32}S)/32 + M(^{56}Fe)/56$, where $N_i$ is the total number of isotope $i$ and $f_i = N_i/\sum_i N_i$.
\item[Composition 3 (C3)~\cite{eichmann2018ultra}] enrichment of the solar composition given by: \\ \hspace*{5cm} $f_i = Z_i f_i^{solar}/ \sum_j Z_j f_j^{solar}$.
\item[Composition 4 (C4)~\cite{aloisio2014ultra}] model based on a homogeneous distribution of sources fitted to the energy spectrum and composition data measured by the Pierre Auger Observatory.
\item[Composition 5 (C5)\cite{liu2012excess}] model based on Wolf-Rayet stellar winds with mass proportion of $M_{He}:M_{C}:M_{O}:M_{Si}=0.32:0.39:0.25:0.04$. The numerical fraction is calculated by $N_i = M_i / MA_i$, with $MA$ the atomic mass of the considered atom, and, again for simplicity, $f_H = f_C + f_O$.
\end{description}

The fractions of nuclei injected by the source in each composition scenario are presented in the table~\ref{tab:composition}, together with the mean charge $\Bar{Z}$ of the injection. These composition models were chosen to cover a wide range of possibilities, from very light to very heavy scenarios.

\begin{table}
\begin{center}
\begin{tabular}{ |c||c|c|c|c|c||c|}
  \hline
  & $f_{H}$ & $f_{He}$ & $f_{N}$ & $f_{Si}$ & $f_{Fe}$ & $\Bar{Z}/e$\\ \hline \hline
  C1 & 0.922 & 0.078 & 8$\times10^{-4}$ & 8$\times10^{-5}$ & 3$\times10^{-5}$ & 1.085 \\ \hline
  C2 & 0.916 & 0.083 & 4.2$\times10^{-4}$ & 5.7$\times10^{-5}$ & 1.5$\times10^{-5}$ & 1.086 \\ \hline
  C3 & 0.849 & 0.1437 & 0.0052 & 0.0010 & 7.2$\times10^{-4}$ & 1.206 \\ \hline
  C4 & 0.7692 & 0.1538 & 0.0461 & 0.0231 & 0.00759 & 1.920 \\ \hline
  C5 & 0 & 0.62 & 0.37 & 0.01 & 0 & 3.97 \\ \hline \hline
\end{tabular}
\end{center}
\caption{Employed compositions.}
\label{tab:composition}
\end{table}

\subsection{Galactic and Extragalactic Magnetic Field}

The galactic magnetic field is known to have a large effect on the arrival direction of UHECR~\cite{Farrar_2013}. However, our galaxy is much smaller than the distance from Earth to Cen A, therefore, the large majority of secondary neutrinos and photons produced by UHECR on the way from Cen A are going to be produced outside the Milky Way. Because they are chargeless and are produced outside our Galaxy, the arrival direction and flux of secondary neutrinos and photons are independent of the galactic magnetic field.

The extragalactic magnetic field (EGMF) deviates the trajectory of charged cosmic rays. The secondary neutrinos and photons generated in the interactions of the cosmic rays with the photon fields are produced in the direction of propagation of the charged cosmic ray therefore not all secondary neutrinos and photons point straight back to the source. On average, the direction of the secondary neutrinos and photons are deviated from the source position by an amount proportional to the path between the source and the production point of the secondary neutrino and photon (scattering should be taken into account.). This is the path traveled by the charged cosmic ray before producing the secondary neutrino or photon.

The total effect of EGMF in the trajectory of the cosmic rays depends on the magnetic field strenght and structure. Heavier nuclei can be largely deflected faking an almost isotropic sky even when a dominant nearby source is considered, as discusses in references~\cite{Dolag_2005,Sigl_2004,Sigl_2003,Das_2008,Kotera_2009}. The total effect in the secondary neutrino and photon point spread function is expected to be much smaller than the effect calculated for the charged particles. Nevertheless, the possibility to identify the source and the acceleration sites in the sources requires the treatment of this effect.

\begin{figure}[]
  \centering
  \includegraphics[width=1.0\columnwidth]{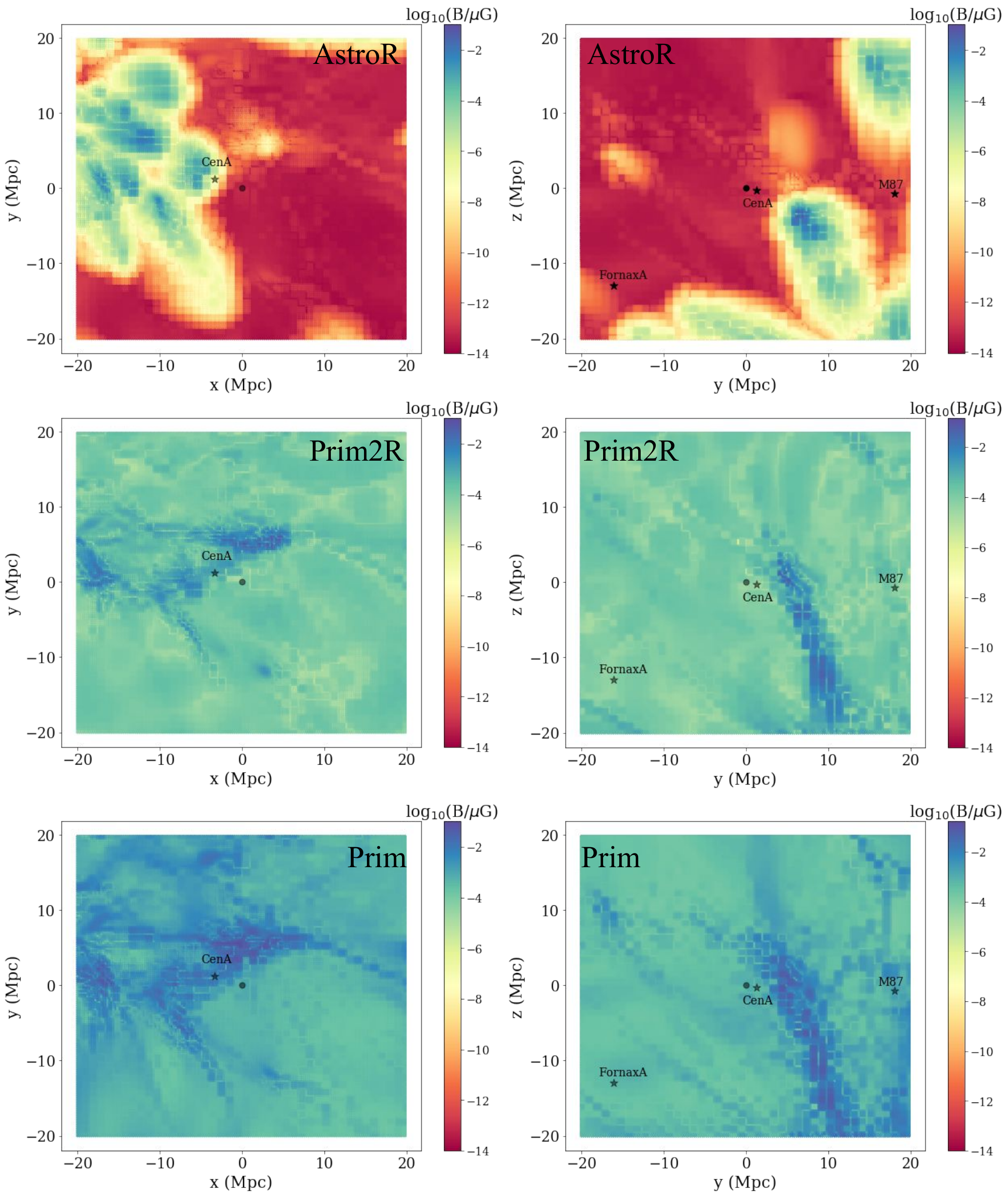}
  \caption{Extragalactic magnetic field intensity (xy- and yz-plane) in an 20 Mpc sphere around Earth for the three models used in this paper: AstroR, Prim2R and Prim according to reference~\cite{hackstein2018simulations}. Stars show the position of Cen A, Fornax A and M87. The center of the map showed by a circles is the the Milky Way.}
  \label{fig:EGMF}
\end{figure}

The EGMF models in reference~\cite{hackstein2018simulations} are used here. These modern computer simulations calculated the EGMF considering its evolution in an expanding universe imposing the current matter distribution as a constraint. Six models for the EGMF with different seed fields are able to describe the constrains. In this paper, three models were chosen to cover the widest range of field intensity within the six models. In reference~\cite{hackstein2018simulations} they were named Primordial (Prim), Primordial2R (Prim2R) and AstrophysicalR (AstroR). Prim and Prim2R models were constructed based on an initial magnetic field at redshift $z=60$ uniform of strength 0.1 nG and a power-law distribution of spectral index -3, respectively. The AstroR was selected as one astrophysical origin EGMF exemplary, were the seed field is developed from cooling and AGN feedback. Figure~\ref{fig:EGMF} shows the field intensity (xy and yz-plane) in an 20 Mpc sphere around Earth.

\subsection{Energy spectrum at the source}

First-order Fermi acceleration is the paradigm for particle acceleration~\cite{longair, kotera2011astrophysics}. Diffusive shock acceleration has been proposed as the main acceleration mechanism in AGN jets~\cite{matthews2019ultrahigh}. Shock waves in the backflow of the jets have been measured in Cen A~\cite{neumayer2007central, hamer2015muse} in which particle acceleration is expected to occur. The energy spectrum can be written as a power law combined with a charge dependent exponential cutoff~\cite{supanitsky2013upper} that accounts for the limitation in the maximal energy that a source can accelerate particles

\begin{equation}
  \frac{dN}{dE dt}= L_{CR} E^{-\gamma} \exp(-E/ZR_{cut}),
  \label{eq:spectrum}
\end{equation}

\noindent where $L_{CR}$ is the cosmic-ray luminosity, $\gamma$ is the spectral index, $Z$ is the charge of the particle and $R_{cut}$ is the rigidity cutoff such as that $E_{cut} = ZR_{cut}$. Continuous emission by the source was considered.

The cosmic-ray luminosity and rigidity cutoff of Cen A was estimated based on the approach developed in reference~\cite{eichmann2018ultra}. The fundamental hypothesis, in agreement with Fermi mechanism, is that $L_{cr}$ must be a fraction of the total kinetic energy in the jet, $Q_{jet}$, which was shown to be proportional to the radio luminosity~\cite{10.1046/j.1365-8711.1999.02907.x} $L_\nu$ at frequency $\nu = 1.1$ GHz ($L_{1.1}$)

\begin{equation}
  L_{cr} = g_{cr} \times A \bigg( \frac{L_{1.1}}{10^{40} \text{erg/s}} \bigg)^{\beta_L} \text{erg/s}
  \label{eq:luminosity}
\end{equation}

\noindent the values of $A$ and $\beta_L$ can be obtained indirectly from data or from theoretical models~\cite{godfrey2016mutual}. Reference~\cite{eichmann2018ultra} uses values obtained for FRII radio galaxies. However, Cen A is classified as a FRI and the values of $\beta_L$ can be different for FRI and FRII~\cite{godfrey2016mutual}. Reference~\cite{cavagnolo2010} used the cavity method, in which the kinetic power of the jet is determined by the radio cavity volume, to calculate $A = 8.6\times 10^{43}$ and $\beta_L = 0.75$ for FRI-type radio galaxies. The measured radio luminosity at 1.1 GHz of Cen A can be found in the Van Velzen's catalog~\cite{van2012radio} $L_{1.1} = 2.6 \times 10^{40}$ erg/s. Therefore equation~\ref{eq:luminosity} becomes $L_{cr} = g_{cr} \times 1.76\times 10^{43}$ erg/s in accordance with references~\cite{dermer2009ultra, yang2012deep}. In this model, $g_{cr}$ contains the relation between the energy stored in hadrons and in the magnetic field. If a source satisfies the minimum energy condition, reference~\cite{pacholczyk1970radio} proposed $g_{cr} = 4/7$. In this paper, instead of fixing the value of $g_{cr}$, it will be fitted to the energy spectrum measured by the Pierre Auger Observatory as explained below.

The rigidity cutoff is related to the escape time from the source. Reference~\cite{eichmann2018ultra} relates $R_{cut}$ to  $L_{1.1}$

\begin{equation}
  R_{cut} = 15 g_{ac} \sqrt{1-g_{cr}} \bigg( \frac{L_{1.1}}{10^{40}\text{ erg/s}} \bigg)^{3/8} \text{EV},
\end{equation}

\noindent $g_{ac}$ depends on plasma physics details in the acceleration site and, is given by (assuming shock acceleration) $g_{ac} = \sqrt{\frac{8\beta_{sh}^2}{f^2\beta_j}}$, with $\beta_{sh}$ the typical shock velocity responsible for the particle acceleration (in speed of light units), $\beta_j$ the jet velocity, and $f$ provide specific plasma properties, with $1 \leq f \leq 8$ for shocks with typical geometries and turbulent magnetic fields~\cite{eichmann2019high}. If Cen A has a jet with velocity $v_j \sim 0.5 c$~\cite{hardcastle2003radio} and a typical velocity of the shock waves $v_{sh} \sim 0.2 c$~\cite{matthews2019ultrahigh}: $0.1 \leq g_{ac} \leq 0.8$.

For each particle type $Z$, the energy spectrum of Cen A used here is determined as a function of three parameters: $\gamma$, $g_{cr}$ and $g_{ac}$. For each of the extragalactic magnetic fields models and composition abundances described in the previous sections, $\gamma$, $g_{cr}$ and $g_{ac}$ are going to be fitted to the energy spectrum of cosmic rays measured by the Pierre Auger Observatory.

\subsection{Energy spectrum at Earth}
\label{sec:fit}

The energy spectrum and composition leaving the source is modified by the interactions with the photon and magnetic fields. These interactions, producing secondary neutrinos and photons, were simulated with the CRPropa 3 framework~\cite{batista2016crpropa}. The relevant energy loss mechanisms and interactions were considered in the propagation of the cosmic rays: adiabatic losses, photodisintegration, photopion production, pair production, and nuclei decay. Only adiabatic losses was considered in the propagation of the secondary neutrinos. Secondary photon were allowed to generate electromagnetic cascades, taking in account inverse Compton scattering, pair production, double and triple pair production. The background photon fields considered were the cosmic microwave background, the extragalactic background light~\cite{gilmore2012semi} and radio~\cite{protheroe1996new}. The photon fields have an important effect on the propagation of high energetic cosmic and gamma rays even for close sources. For instance, the pair-production mean free path of a EeV gamma ray is of the order of Mpc, therefore important for propagation from Cen A. However, the differences between different models of extragalactic background light is important mainly to energies below the interest of this paper~\cite{10.1093/mnras/stt684} and for distances larger than the source considered here~\cite{Batista_2019}. The interaction of UHECR with the gas surrounding Cen A can be safely neglect given the baryon content estimated in the filaments of massive galaxy cluster to have density $\approx 10^{-5}$ cm$^{-3}$~\cite{density}, the mean free path for proton-proton interactions in this medium is of the order of 1 Gpc.

Cosmic rays were injected with energies between $10^{18}$ and $10^{21}$ eV and propagated down to $10^{17}$ eV. The calculations include $10^{9}$ events of $^{1}H$, $^{4}He$, $^{14}N$, $^{28}Si$ and $^{56}Fe$ nuclei for each magnetic field configuration. The particles were injected with a spectral index $-1$ to guarantee equal statistical fluctuations at all energies. The energy spectrum shape and composition fractions were introduced as weights according to the procedure explained in reference~\cite{eichmann2018ultra}. The observational radius was considered to be $r_{obs} = 100$ kpc. An observational radius of 10 kpc was also simulated and the conclusion presented here remains the same.

\begin{figure}[]
  \centering
  \includegraphics[width=0.8\columnwidth]{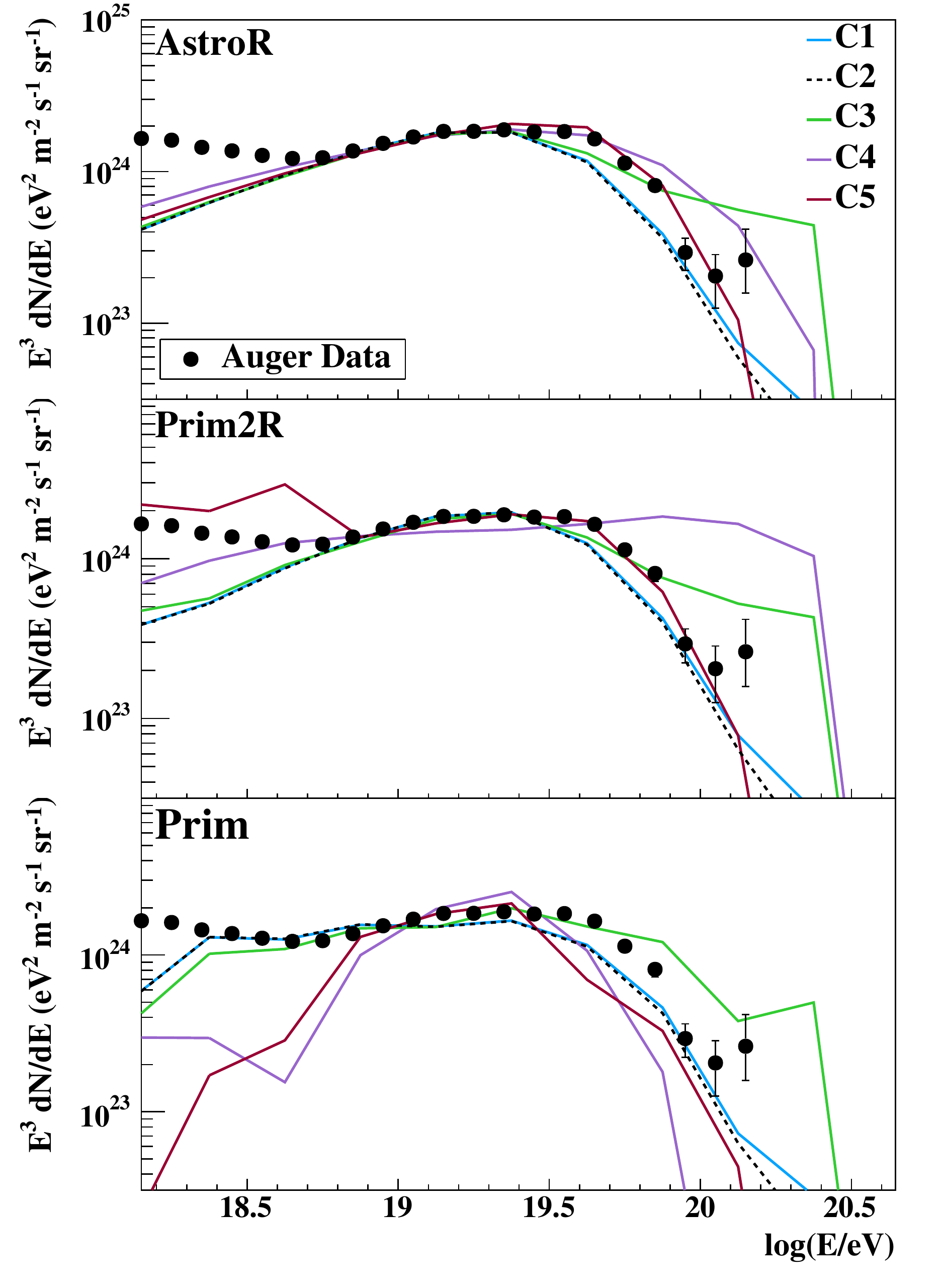}
  \caption{Flux of UHECR as a function of energy. Dots show measurements done by the Pierre Auger Observatory~\cite{aab2020measurement}. Lines show the result of the fit as explained in section~\ref{sec:fit}. Three EGMF models (AstroR, Prim2R and Prim) and five composition abundances (C1--C5) are shown.}
  \label{fig:cr:fit}
\end{figure}

The simulated energy spectrum arriving on Earth has three parameters $\gamma$, $g_{cr}$ and $g_{ac}$ which were fitted to the energy spectrum measured by the Pierre Auger Observatory~\cite{aab2020measurement}. The parameters $\gamma$, $g_{cr}$ and $g_{ac}$ were allowed to vary between $[2.00, 2.30]$, $[0.10, 1.00]$ and $[0.01,1.00]$ in steps of 0.02, respectively. These parameter limits brackets the possible values within the classical model used here, which guarantees the coherent use of the theory. However, it is important to notice that some sophisticated models includes, for instance, harder energy spectrum~\cite{Unger_2015}. In this fit, 100\% of the flux measured by the Pierre Auger Observatory was considered to be generated by Cen A. This is probability a simplification because other sources beyond Cen A are expected to generate a non negligible flux of cosmic rays arriving at Earth. However, this simplification is acceptable because the fit is used only to fine tune the shape of the energy spectrum predicated by acceleration models. By bracketing the limits of the parameters with the theoretical prediction, no inconsistency is propagated to the further analysis. The contribution of Cen A to the total flux of cosmic rays will be allowed to vary in the next section to remove this simplification when needed. Equation~\ref{eq:spectrum} was fitted to the energy spectrum data using the standard $\chi^2$ method including uncertainties for energies above $10^{18.7}$ eV. Table~\ref{tab:fit} and figure~\ref{fig:cr:fit} shows the results of the fits.

It is beyond the scope of this paper to evaluate how well each astrophysical model describes the data measured by the Pierre Auger Observatory. For that purpose, the fit would have to include other measured quantities (i.e. Xmax distribution) and more free parameters in the models. The fit implemented in this study has a limited focus on improving the calculation of $g_{cr}$ and $g_{ac}$ normalizing the simulations to the flux measured by the Pierre Auger Observatory. The values of these parameters have been estimated based on theoretical assumptions as shown above. In the fit presented here, the theoretical estimations are used as limits and first guesses to search the best values of $g_{cr}$ and $g_{ac}$ that describes the energy spectrum. Note that some models produces a flux higher than the one measured by the Pierre Auger Observatory, which in principal could be used to invalidate the model, however small changes to the models could ameliorate the agreement with the data.

\begin{table}
  \begin{center}
    \begin{tabular}{|c|c|c|c|c|c|c|c|c|c|c|c|c|} \hline
      & \multicolumn{4}{c|}{AstroR} & \multicolumn{4}{|c|}{Prim2R} & \multicolumn{4}{c|}{Prim} \\ \cline{2-13}
      Composition & $\gamma$ & $g_{ac}$ & $g_{cr}$ & $\chi^2/dof$ & $\gamma$ & $g_{ac}$ & $g_{cr}$ & $\chi^2/dof$ & $\gamma$ & $g_{ac}$ & $g_{cr}$ & $\chi^2/dof$ \\ \hline
      C1 & 2.12 & 0.97 & 0.12 & 17.6 & 2.10 & 0.95 & 0.10 & 11.8 & 2.12 & 0.99 & 0.12 & 26.4 \\ \hline
      C2 & 2.12 & 0.97 & 0.12 & 19.0 & 2.10 & 0.95 & 0.10 & 12.6 & 2.12 & 0.99 & 0.12 & 30.0 \\ \hline
      C3 & 2.18 & 0.97 & 0.26 & 11.3 & 2.14 & 0.99 & 0.14 & 8.5 & 2.12 & 0.99 & 0.10 & 12.3 \\ \hline
      C4 & 2.08 & 0.09 & 0.12 & 2.9 & 2.26 & 0.99 & 0.38 & 60.9 & 2.04 & 0.13 & 0.98 & 165.0 \\ \hline
      C5 & 2.28 & 0.73 & 0.82 & 0.6 & 2.26 & 0.59 & 0.86 & 1.8 & 2.28 & 0.33 & 0.58 & 42.4 \\ \hline \hline
    \end{tabular}
  \end{center}
  \caption{Results of the fit of the energy spectrum.}
  \label{tab:fit}
\end{table}

\section{Results I: secondary neutrinos from Cen A}
\label{sec:result:neutrino}

\begin{figure}[]
  \centering
  \includegraphics[width=0.8\columnwidth]{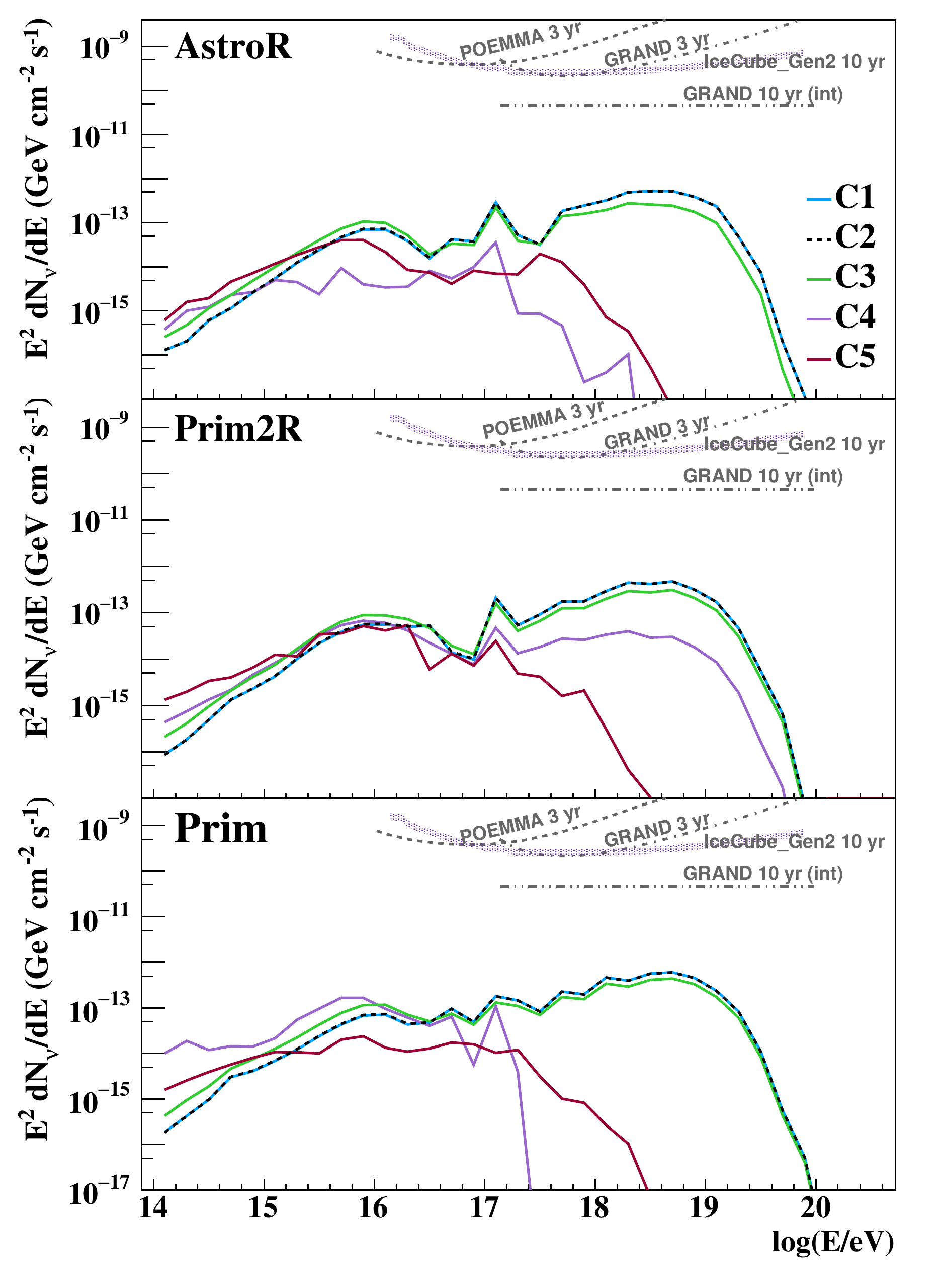}
  \caption{Flux of secondary neutrino as a function of energy. Three EGMF are shown: AstroR, Prim2R and Prim. Each curve corresponds to a composition abundance: C1--C5. The sensitivity of future experiments POEMMA~\cite{venters2019poemma} and GRAND~\cite{grand2018}, IceCube\_2Gen~\cite{aartsen2020icecube} are shown.}
  \label{fig:neutrino:flux}
\end{figure}

This section presents the results about the energy spectrum and arrival direction of secondary neutrinos from Cen A. Figure~\ref{fig:neutrino:flux} shows the flux of secondary neutrinos from Cen A for all astrophysics scenarios considered here. The sensitivity of the futures observatories GRAND~\cite{grand2020}, POEMMA~\cite{venters2019poemma} and IceCube\_2Gen~\cite{aartsen2020icecube} are also shown. The effect of the injection composition is clearly seen. Heavier compositions (C4 and C5) produce less neutrinos at the highest energies because nuclear fragmentation is more likely than photo-pion production. The flux of secondary neutrinos from Cen A is much smaller than the sensitivity of the future experiments for all scenarios tested here.

\begin{figure}[]
  \centering
  \includegraphics[width=0.85\columnwidth]{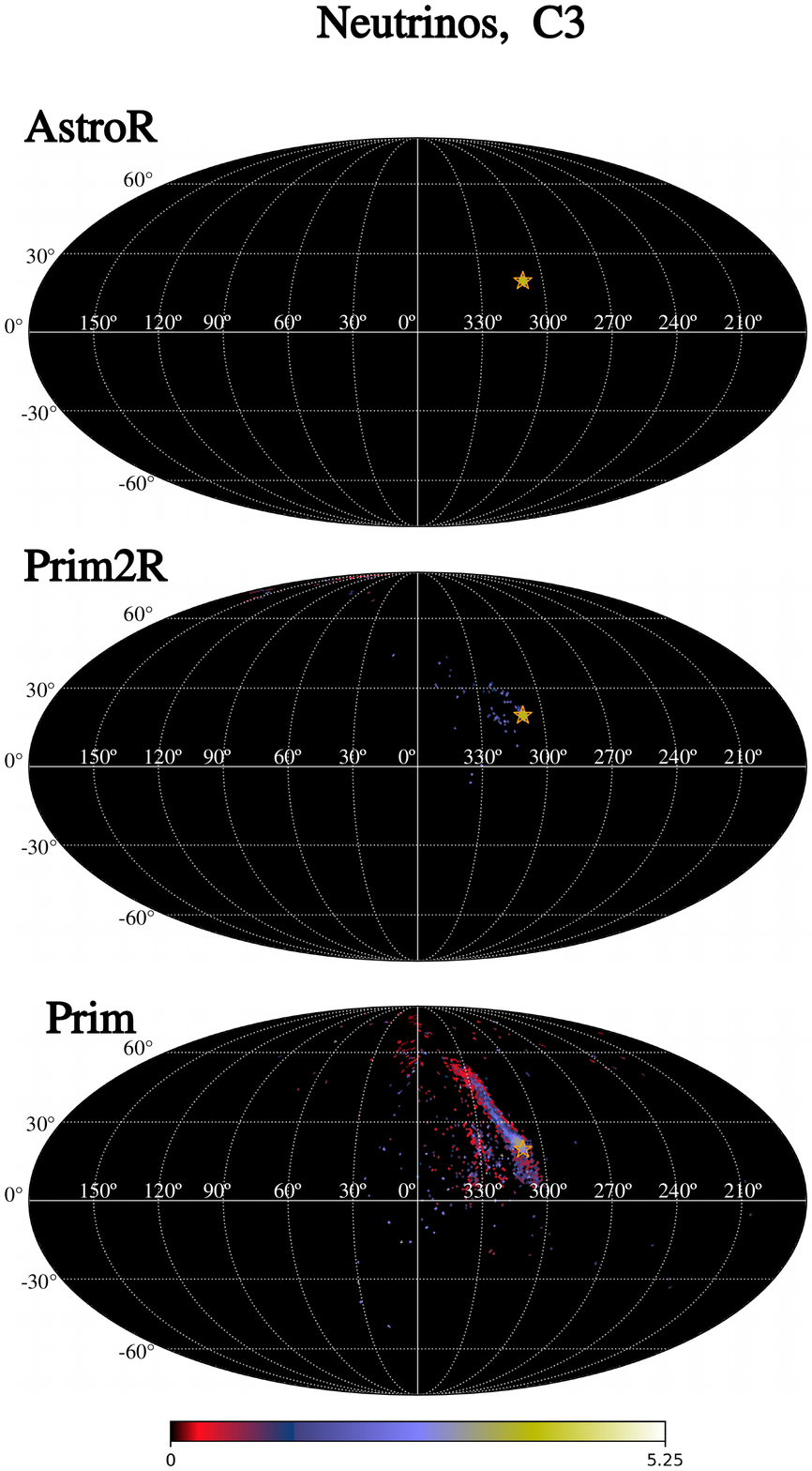}
  \caption{Arrival direction map of secondary neutrinos for the three EGMF models and C3 composition abundance. The star indicates the position of Cen A. The maps are in galactic coordinates and Mollweide projection.}
  \label{fig:neutrino:arrival:map}
\end{figure}

\begin{figure}[]
  \centering
  \includegraphics[width=0.8\columnwidth]{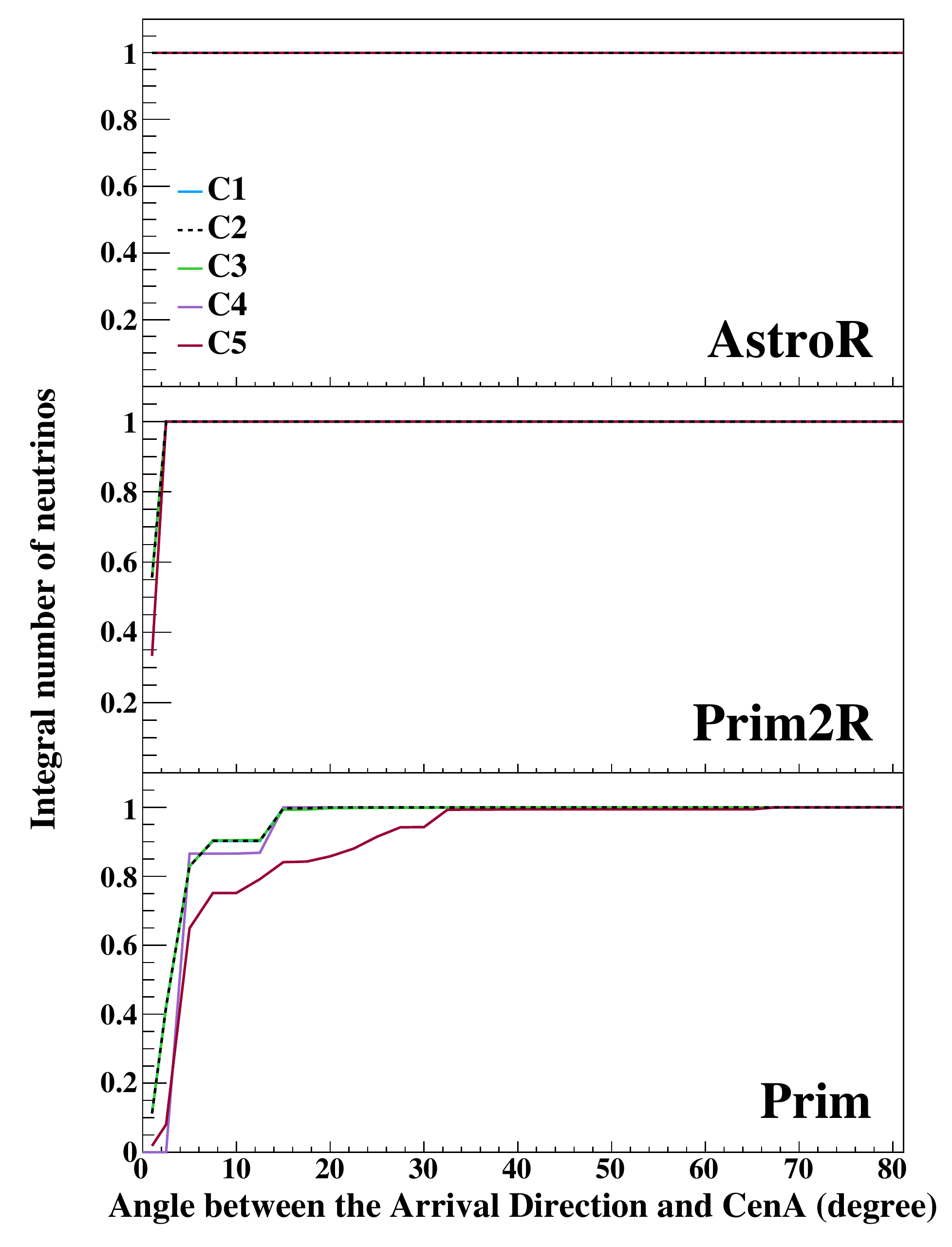}
  \caption{Integral distribution of the angular distance between Cen A and the arrival direction of secondary neutrinos. Three EGMF are shown: AstroR, Prim2R and Prim. Each curve corresponds to a composition abundance: C1--C5.}
  \label{fig:neutrino:arrival:distribution}
\end{figure}

Figure~\ref{fig:neutrino:arrival:map} shows maps of the simulated arrival direction of neutrinos for the three magnetic field models (AstroR, Prim2R and Prim) for the C3 composition abundance. Similar plots were produced for all composition scenarios. Figure~\ref{fig:neutrino:arrival:distribution} shows the cumulative angular difference between the arrival direction of the neutrino and the direction of Cen A. It is clear that arrival direction of neutrinos depends significantly on the extragalactic magnetic field model. If AstroR model is used in the calculations, secondary neutrinos point direct back to Cen A. If Prim2R model is used in the calculations, secondary neutrinos are distributed around a window of $2^\circ$ around Cen A. If Prim model is used in the calculations, secondary neutrinos are distributed in large window, from $14^\circ$ to $32^\circ$, around Cen A.

\begin{figure}[]
  \centering
  \includegraphics[width=0.7\columnwidth]{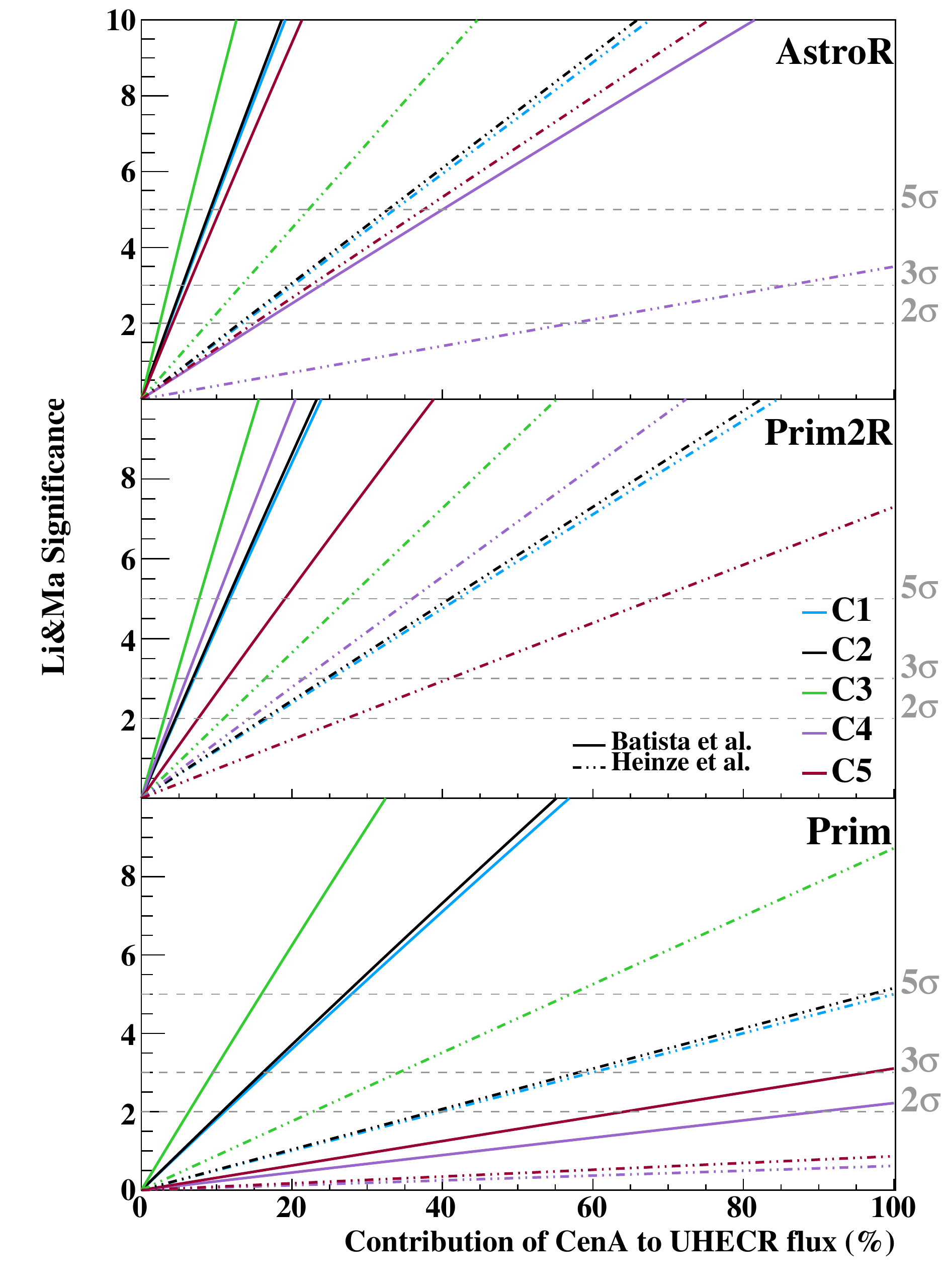}
  \caption{Li-Ma significance of the secondary neutrino signal as a function of the contribution from Cen A to the total UHECR flux measured by the Pierre Auger Observatory for energies above 60 EeV. The significance was calculated supposing 200 000 km$^2$ detection area observatory taking data during one year. Full (Dashed-dotted) lines represents the calculations when cosmogenic background neutrinos were estimated according to Batista et al.~\cite{batista2019cosmogenic} (Heinze et al.~\cite{heinze2019new}) and considered as the null hypothesis. The three EGMF models (AstroR, Prim2R and Prim) and five composition abundance (C1--C5) are shown.}
  \label{fig:neutrino:lima}
\end{figure}

The possibility of detection of an anisotropic signal of secondary neutrinos from Cen A was calculated using the Li-Ma significance~\cite{li1983analysis,lima:Neutrons_auger}. The background given by cosmogenic neutrino flux, null hypothesis in the statistical test, was taken from references~\cite{batista2019cosmogenic} and~\cite{heinze2019new}. Figure~\ref{fig:neutrino:lima} shows the significance obtained for neutrinos with $E >10^{15}$ eV inside a circular window of radius $2.5^\circ$ centered in Cen A for the two cosmogenic neutrino models considering one year data of the GRAND observatory (200 000 km$^2$ detection area~\cite{grand2020}). The significance is shown as a function of the contribution of Cen A to the total flux of cosmic rays measured by the Pierre Auger Observatory. Large anisotropic signals ($> 5\sigma$) could be detected for the AstroR and Prim2R EGMF models even if Cen A contributed with a small percentage of the total flux of cosmic rays irrespective of the cosmogenic neutrino model used in the calculation. For the Prim EGMF model, the anisotropic signal detected is highly composition dependent.

\section{Results II: Secondary photons from Cen A}
\label{sec:result:photon}

\begin{figure}[]
  \centering
  \includegraphics[width=0.9\columnwidth]{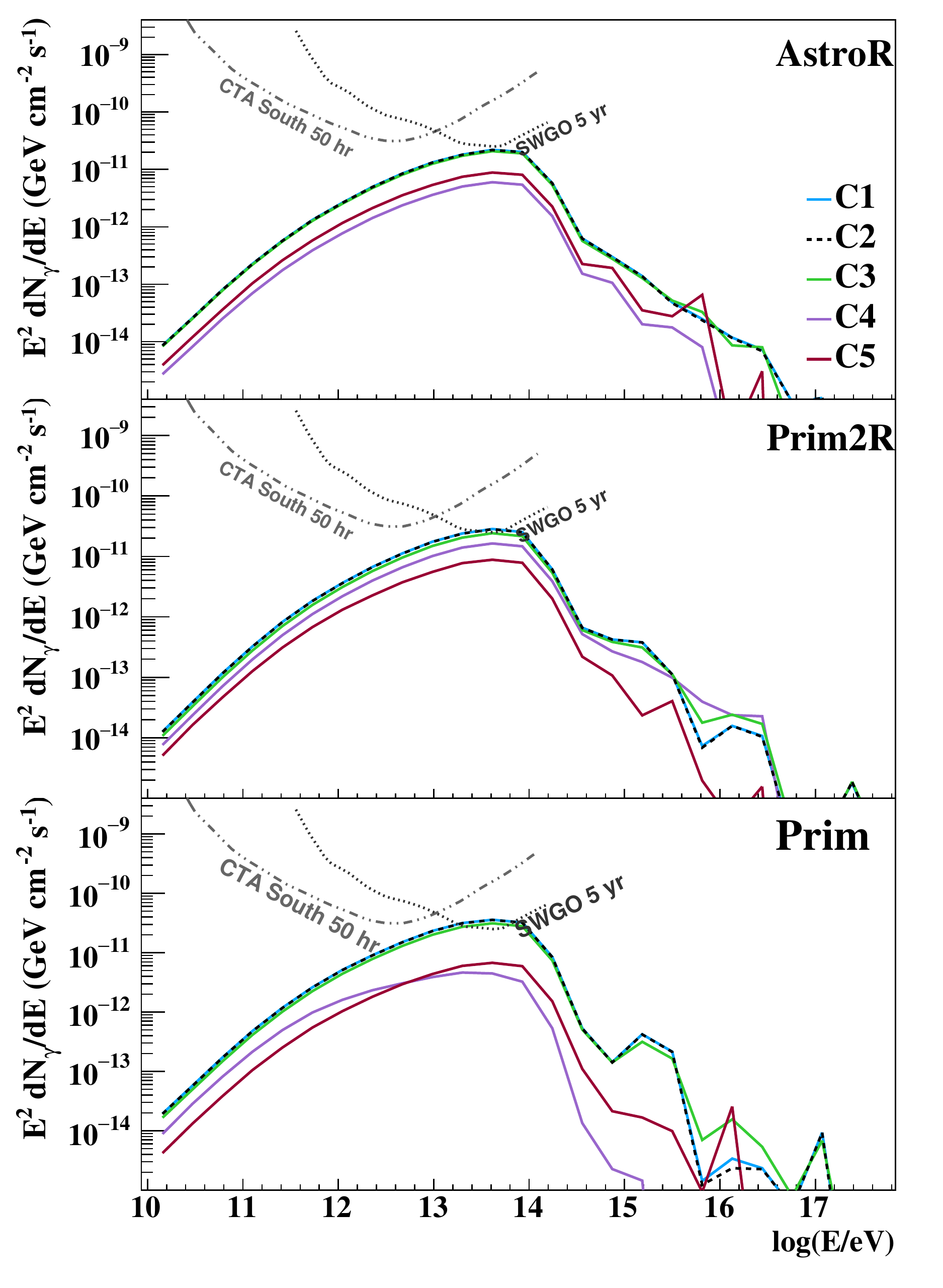}
  \caption{Flux of secondary gamma-rays as a function of energy. Three EGMF are shown: AstroR, Prim2R and Prim. Each curve corresponds to a composition abundance: C1--C5. The sensitivity of future experiments SWGO~\cite{swgo} and CTA~\cite{hassan2017monte} are shown.}
  \label{fig:photon:flux}
\end{figure}

\begin{figure}[]
  \centering
  \includegraphics[width=0.8\columnwidth]{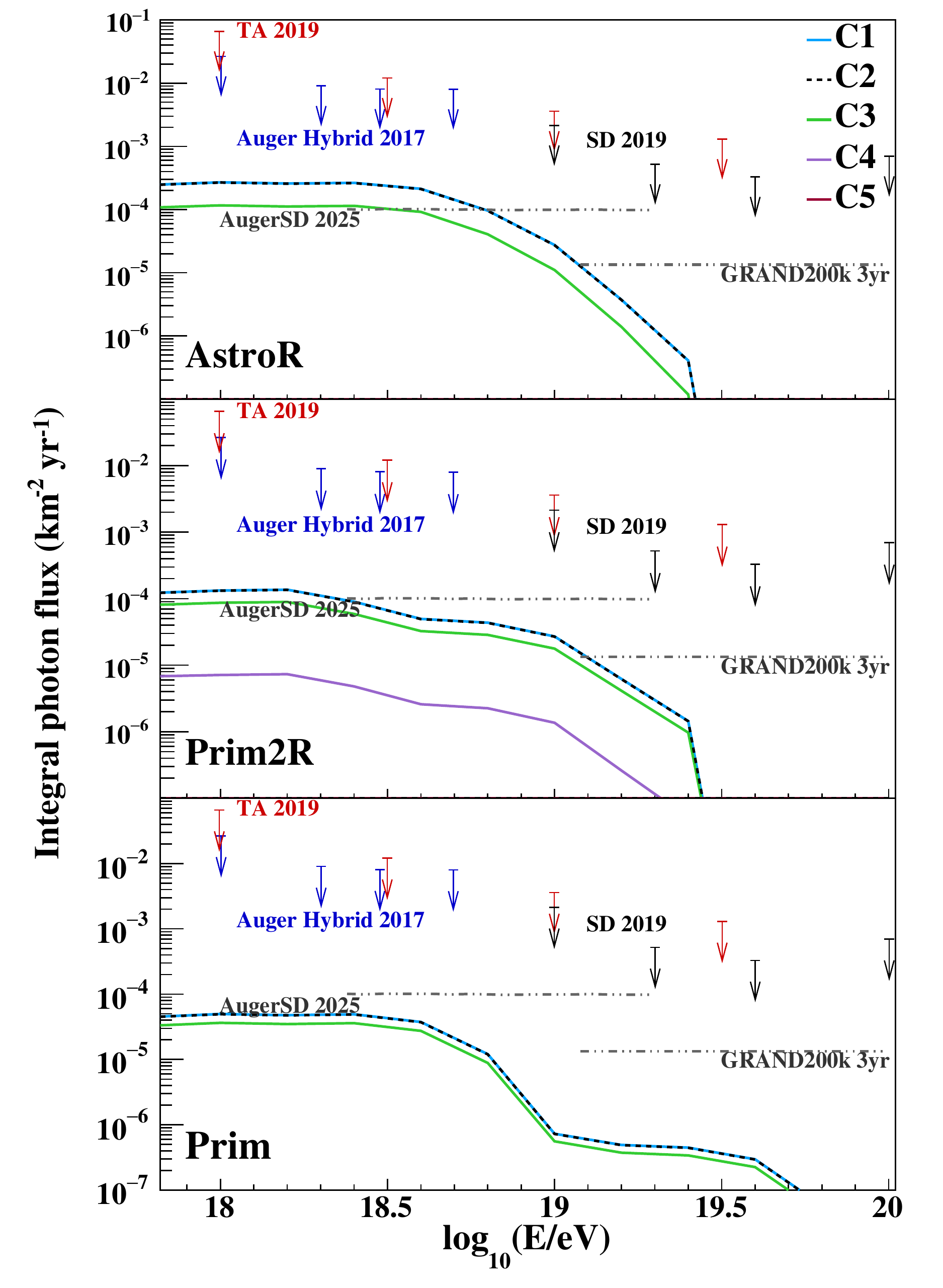}
  \caption{Integral photon flux and limits in the highest energy limit. The three EGMF models (AstroR, Prim2R and Prim) and five composition abundance (C1--C5) are shown. Arrows show the upper limits measured by Telescope Array (TA 2019)~\cite{TA_phot_2019} and by the Pierre Auger Observatory (Auger Hybrid 2017~\cite{aab2017search} and SD 2019~\cite{rautenberg2019limits}). The projected sensitivity for the Pierre Auger Observatory in 2025 (AugerSD 2025~\cite{rautenberg2019limits}) and for GRAND (GRAND200K 3yr~\cite{grand2018}) are also shown.}
  \label{fig:photon:integral:flux}
\end{figure}

The same calculations regarding flux and arrival direction were done for the secondary photons produced by UHECR on the way from Cen A to Earth. Figure~\ref{fig:photon:flux} shows the flux of secondary photons arriving at Earth in comparison to the sensitivity of current and future gamma-ray observatories. The SWGO~\cite{swgo} experiment will have the highest probability to detect a signal because its maximum sensitivity is at the energy range (from 0.5 to 1$\times 10^{14}$ eV) in which the highest flux of photons arrives on Earth. Figure~\ref{fig:photon:integral:flux} shows the calculated integral photon flux of secondary photons in comparison to current measurements~\cite{TA_phot_2019,aab2017search,rautenberg2019limits} and future predictions~\cite{grand2018}. If compositions similar to C1, C2 and C3 models are emitted by Cen A and the magnetic field is not so extreme as Prim model, the Pierre Auger Observatory will measure a photon signal from Cen A by 2025 and the GRAND experiment will confirm this measurement after three years of operation.

\begin{figure}[]
  \centering
  \includegraphics[width=0.85\columnwidth]{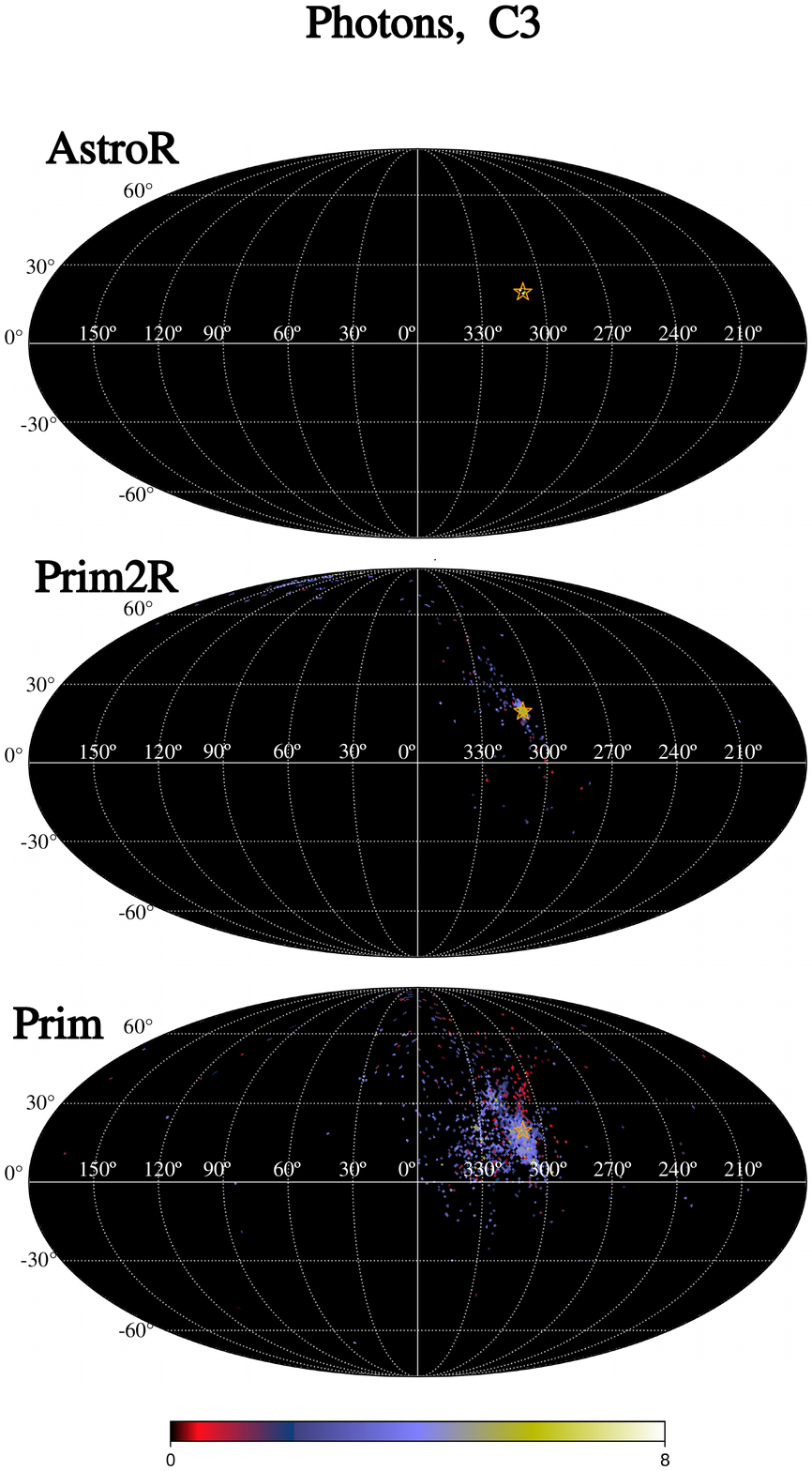}
  \caption{Arrival directions map of secondary photons for the three EGMF models and C3 composition abundance. The star indicates the position of Cen A. The maps are in galactic coordinates and Mollweide projection.}
  \label{fig:photon:arrival:map}
\end{figure}

\begin{figure}[]
  \centering
  \includegraphics[width=0.9\columnwidth]{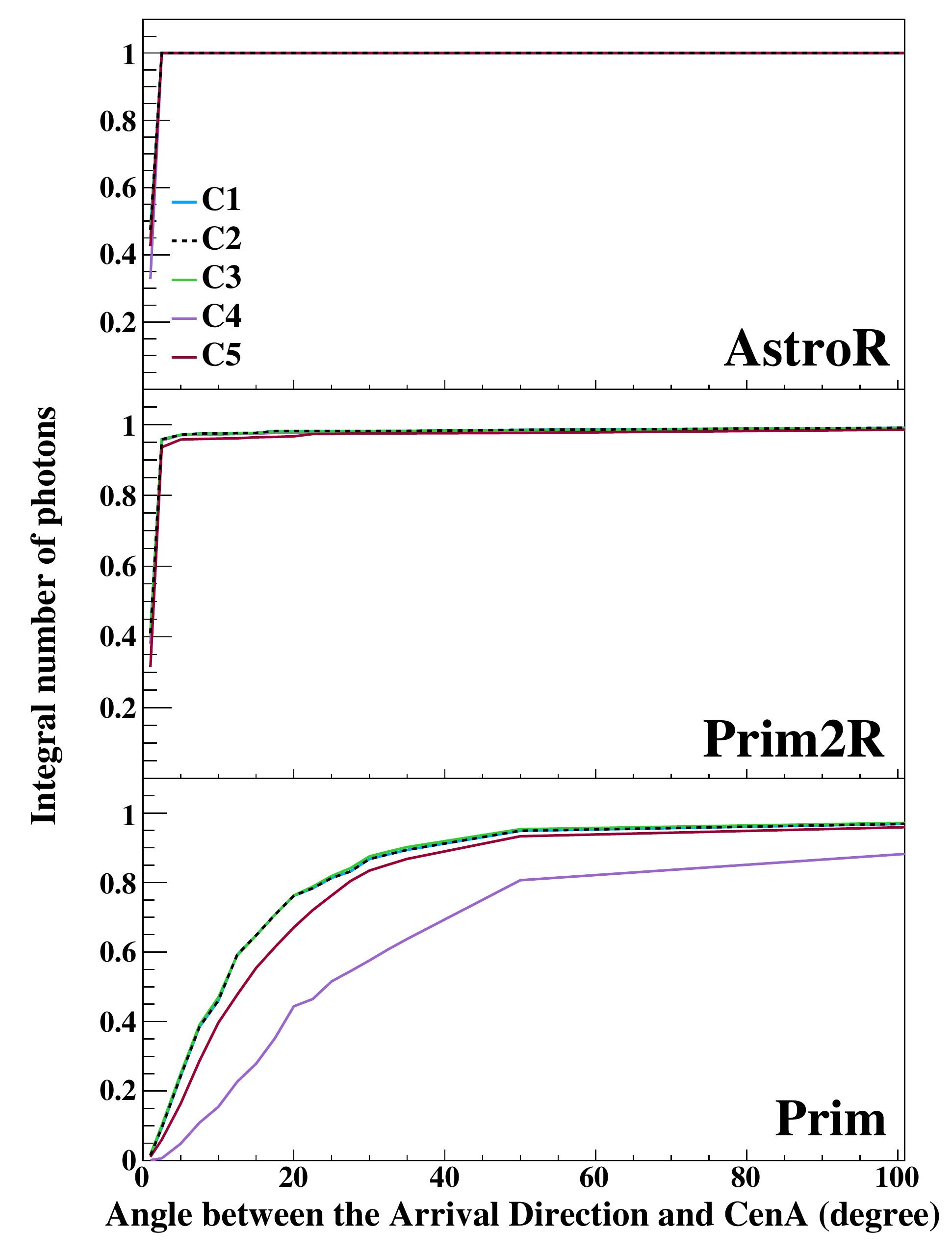}
  \caption{Integral distribution of the angular distance between Cen A and the arrival direction of secondary photons with E$<10^{17}$ eV. Three EGMF are shown: AstroR, Prim2R and Prim. Each curve corresponds to a composition abundance: C1--C5.}
  \label{fig:photon:arrival:integral}
\end{figure}

Figure~\ref{fig:photon:arrival:map} shows the sky map of photons for the C3 composition model as an example. Similar plots were produced for all composition models. Figures~\ref{fig:photon:arrival:integral} show the integral distribution of the angular distance from Cen A. The Prim model generates a large deviation of the UHECR resulting in a distribution of photons pointing back to directions far away from Cen A.

\begin{figure}[]
  \centering
  \includegraphics[width=0.75\columnwidth]{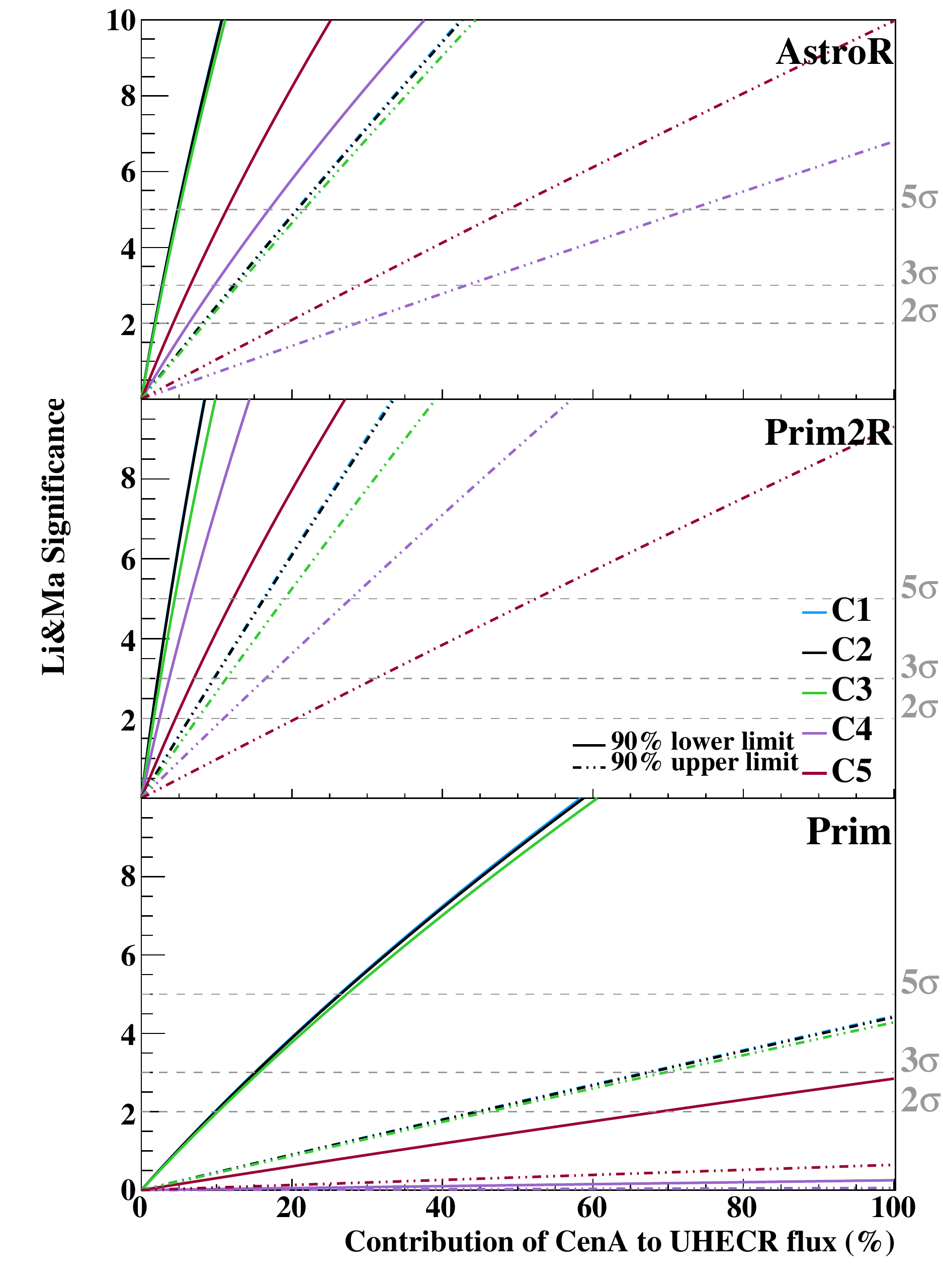}
  \caption{Li-Ma significance of the photon signal as a function of the contribution from Cen A to the total UHECR flux measured by the Pierre Auger Observatory for energies above 60 EeV. The significance was calculated supposing 80 000 m$^2$ detection area observatory taking data during five years. Full (Dashed-dotted) lines represents the calculations when the 90\% lower (upper) limit flux of cosmogenic background photons estimated in reference~\cite{batista2019cosmogenic} was considered as the null hypothesis. The three EGMF models (AstroR, Prim2R and Prim) and five composition abundance (C1--C5) are shown.}
  \label{fig:photon:lima}
\end{figure}

The Li-Ma significance of the photon signal around Cen A was also calculated. The 90\% upper and lower limits of cosmogenic photon flux predicted by reference~\cite{batista2019cosmogenic} was used as the null hypothesis in the statistical test. The significance obtained for photons with $E >10^{12}$ eV inside a circular window of radius $2.5^\circ$ centered in Cen A is shown in figure~\ref{fig:photon:lima}. These results were evaluated to $\sim 10^{17}$ cm$^2$ s, or five year of measurement by the SWGO observatory with 80 000 m$^2$ detection area~\cite{swgo}. Large anisotropic signals ($> 5\sigma$) could be detected for the AstroR and Prim2R EGMF models even if Cen A contributed with a intermediate fraction of the total flux of cosmic rays. For the Prim EGMF model, the photon signal is sufficient blurred, making the detection unlikely if Cen A contribute with less than 60\% to the UHECR flux.

\section{Conclusions}
\label{sec:conclusion}

Intrinsic TeV photons from Cen A have been measured and intrinsic neutrinos could be detected by future experiments. UHECR produced in Cen A would produce secondary neutrinos and photons on the way to Earth. The detection of high energy secondary photons and neutrinos would add another layer of understanding in the UHECR production because they are produced in a medium and by mechanisms better understood than the ones happening inside the source.

Fifteen astrophysical models representing the combination of three EGMF and five composition abundances were studied. The flux and arrival direction of secondary neutrinos and photons were calculated. The possibility of current and future observatories to measure the flux of secondary neutrino and flux was investigated. The possibility to detected an anisotropic signal of secondary neutrinos and photons in the direction of Cen A was also calculated. The results depends strongly on the  extragalactic magnetic field model, do not depend on the galactic field and depends at some level on the composition abundance. The effect of the magnetic field in the deflection of the UHECR before the production of the photon and neutrino is large enough in some models to erase the signal.

It was shown here for the first time how important is the local extragalactic magnetic field in the evaluation of the flux and arrival direction of neutral secondary particles. The effect of the extragalactic magnetic field on the arrival direction and mainly flux has been widely neglected in the literature. Simple models of extragalactic magnetic field have been used in previous studies but the results presented here show that the struture of the local magnetic field should be taken into account even when secondary neutral particles are studied. The Prim EGMF model erases any signal of anisotropy of secondary photons and neutrinos from Cen A. The effect of the local extragalactic magnetic field in the detection of charged particles from local sources will be presented in a future publication.

The secondary neutrinos from Cen A could be detected as an anisotropic signal in the arrival direction with statistical significance by planned experiments such as GRAND, IceCube\_Gen2 and POEMMA if the local extragalactic magnetic fields is described by AstroR or Prim2R models.

The secondary photons from Cen A could be detected by the Pierre Auger Observatory as a excess above $10^{18.5}$ eV in the integral photon flux by 2025 if the emmited composition is light such as models C1, C3 e C3 independently of the structure of the local extragalactic field. This detection could be confirmed by three years of operation of the GRAND experiment. The secondary photons from Cen A could also be detected as an anisotropic signal in the arrival direction with statistical significance by planned experiments such as SWGO if the local extragalactic magnetic fields is described by AstroR or Prim2R models.

The detection of secondary neutrinos or photons from Cen A would unequivocally identify a UHECR source and start a new era in cosmic rays physics.

\section*{Acknowledgments}

CO and VdS acknowledge FAPESP Project 2015/15897-1 and 2018/24256-8. Authors acknowledge the National Laboratory for Scientific Computing (LNCC/MCTI,  Brazil) for providing HPC resources of the SDumont supercomputer (http://sdumont.lncc.br). VdS acknowledges CNPq.

\bibliographystyle{elsarticle-num}
\bibliography{ref}

\appendix
\section{Intrinsic neutrino flux normalization}
\label{app:flux}

\begin{figure}[]
  \centering
  \includegraphics[width=1.0\columnwidth]{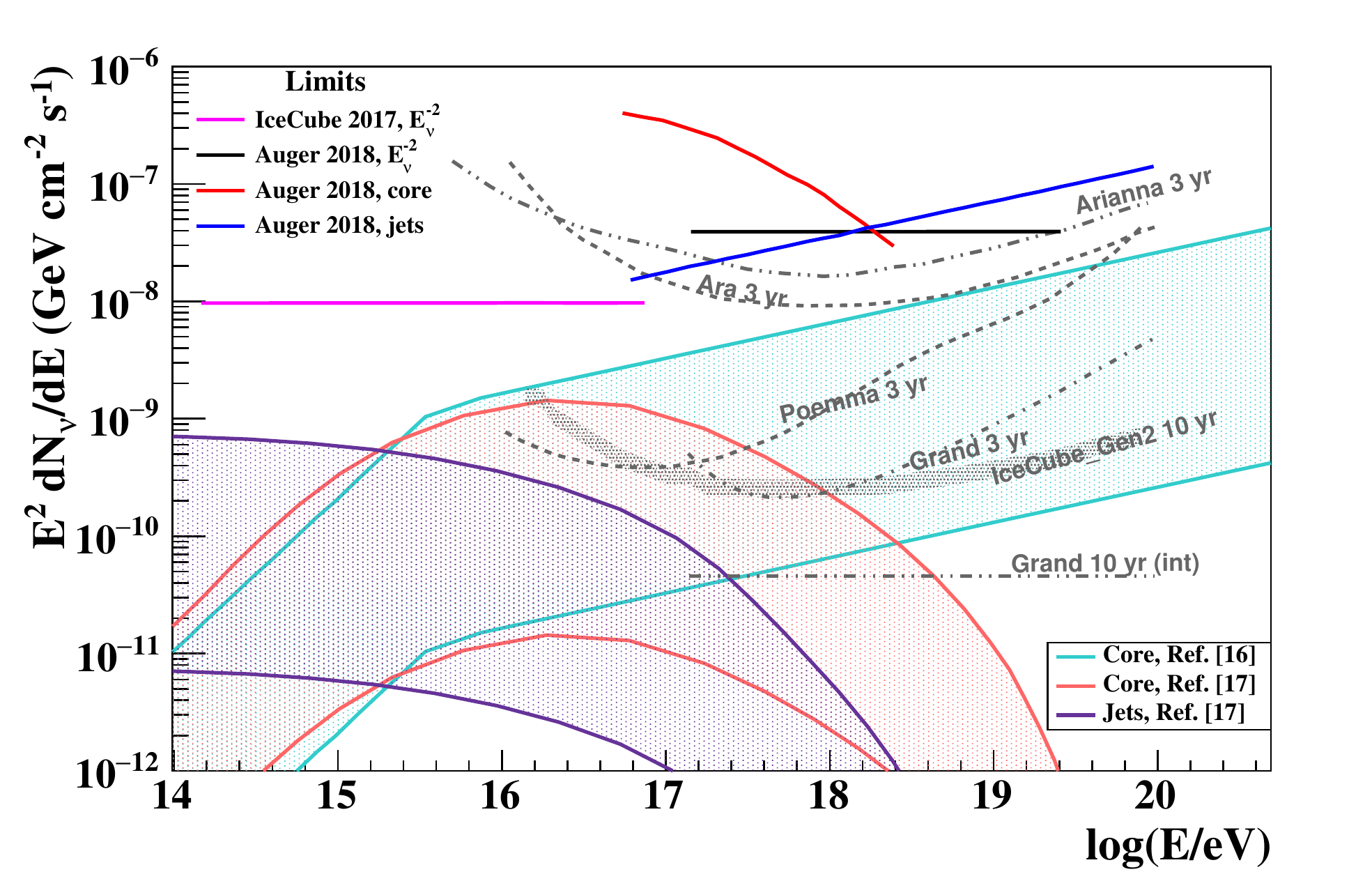}
  \caption{Flux of intrinsic neutrino produced in Cen A according to reference~\cite{cuoco2008ultrahigh, kachelriess2009high} (hatched regions) as function of the contribution of Cen A to the total flux of UHECR measured by the Pierre Auger Observatory with energy above $E_c = 60$ EeV~\cite{Verzi:2019AO} and the normalization given in equation~\ref{eq:norma}. Upper limits on the neutrino flux measured by the Pierre Auger Observatory~\cite{aab2019limits} and the IceCube experiment \cite{aartsen2017all} are shown as full lines. Dashed and dashed-dotted lines shows the sensitivity of future experiments not corrected to the direction of Cen A.}
  \label{fig:neutrino:intrinsic}
\end{figure}

 The flux normalization in figure~\ref{fig:neutrino:intrinsic} was done taking into account all UHECR measured irrespective of the direction. The total intrinsic neutrino flux coming from Cen A can be recalculated as

\begin{equation}
  F_\nu^{2018}(E) = \xi \times F_{\nu}^{2008}(E) \times \frac{\Xi^{2008}}{\Xi^{2018}} \times \frac{N_{ev}^{2018}}{N_{ev}^{2008}}\Big|_{E\geq E_c} \; \; ,
  \label{eq:norma}
\end{equation}

\noindent where $E_c = 60$ EeV, $N_{ev}^{2018} = 122$ is the total number of events detected by the Pierre Auger Observatory with energy $E>E_c$~\cite{Verzi:2019AO} and $N_{ev}^{2008} = 2$ is the number of events attributed to Cen A in the original proposal of the models. $\Xi$ the exposure of the Pierre Auger Observatory in the years considered, $\Xi^{2018} = 60400$ km$^2$ sr yr~\cite{Verzi:2019AO} and $\Xi_{2008} = 9000$ km$^2$ sr yr~\cite{abraham2008astropart}. $\xi$ is a factor to convert from the point source coverage to full sky coverage. $\xi = 0.11$ for the window considered in reference~\cite{kachelriess2009high} and $\xi = 0.20$ for the window considered in reference~\cite{cuoco2008ultrahigh}.

%===============================================
%FIGURES

% intrinsic neutrino

% Astro model
%===============================================

%===============================================
%NEUTRINOS

%==============================
% photon

\end{document}